\documentclass[twocolumn,fleqn,usenatbib,useAMS]{mnras}

\usepackage[T1]{fontenc}
\usepackage{ae,aecompl}
\usepackage{hyperref}
\usepackage{soul}
\usepackage{xcolor}

\DeclareRobustCommand{\VAN}[3]{#2}
\let\VANthebibliography\thebibliography
\def\thebibliography{\DeclareRobustCommand{\VAN}[3]{##3}\VANthebibliography}
\usepackage{graphicx}	
\usepackage{amsmath}	
\usepackage{amssymb}	
\usepackage{multicol}        
\usepackage{bm}		
\usepackage{pdflscape}	

\title[Hot Accretion Flow with Thermal Conduction]{Global Transonic Solution of Hot Accretion Flow with Thermal Conduction}
\author[Mitra et al.]{
	Samik Mitra$^{1}$\thanks{m.samik@iitg.ac.in (SM)},
	Sayyedeh Masoumeh Ghoreyshi$^{2}$\thanks{smghoreyshi64@gmail.com (SMG, corresponding author)},
	Amin Mosallanezhad$^{3}$\thanks{mosallanezhad@xjtu.edu.cn (AM)},
	\newauthor
	Shahram Abbassi$^{2}$,	and Santabrata Das$^{1}$\thanks{sbdas@iitg.ac.in (SD)}
	\\
	$^{1}$Indian Institute of Technology Guwahati, Guwahati 781039, Assam, India\\
	$^{2}$Department of Physics, School of Science, Ferdowsi University of Mashhad, Mashhad, PO Box 91775-1436, Iran\\
    $^{3}$School of Mathematics and Statistics, Xi'an Jiaotong University, Xi'an, Shaanxi 710049, China
}

\date{Accepted XXX. Received YYY; in original form ZZZ}

\pubyear{2023}

\begin{document}
	\label{firstpage}
	\pagerange{\pageref{firstpage}--\pageref{lastpage}}
	\maketitle
	
\begin{abstract}
	
	We examine the effect of thermal conduction on the low-angular momentum hot accretion flow (HAF) around non-rotating black holes accreting mass at very low rate. While doing so, we adopt the conductive heat flux in the saturated form, and solve the set of dynamical equations corresponding to a steady, axisymmetric, viscous, advective accretion flow using numerical methods. We study the dynamical and thermodynamical properties of accreting matter in terms of the input parameters, namely energy ($\varepsilon_0$), angular momentum ($\ell_0$), viscosity parameter ($\alpha$), and saturation constant ($\Phi_{\rm s}$) regulating the effect of thermal conduction. We find that $\Phi_{\rm s}$ plays a pivotal role in deciding the transonic properties of the global accretion solutions. In general, when $\Phi_{\rm s}$ is increased, the critical point ($r_{\rm c}$) is receded away from the black hole, and flow variables are altered particularly in the outer part of the disc. To quantify the physically acceptable range of $\Phi_{\rm s}$, we compare the global transonic solutions with the self-similar solutions, and observe that the maximum saturation constant ($\Phi^{\rm max}_{\rm s}$) estimated from the global solutions exceeds the saturated thermal conduction limit ($\Phi_{\rm sc}$) derived from the self-similar formalism. Moreover, we calculate the correlation between $\alpha$ and $\Phi^{\rm max}_{\rm s}$ and find ample disagreement between global solutions and self-similar solutions. Further, using the global flow variables, we compute the Bernoulli parameter ($Be$) which remains positive all throughout the disc, although flow becomes loosely unbound for higher $\Phi_{\rm s}$. Finally, we indicate the relevance of this work in the astrophysical context in explaining the possibility of massloss/outflows from the unbound disc.
\end{abstract}
	
	\begin{keywords}
		accretion, accretion disc -- black hole physics -- conduction -- hydrodynamics
	\end{keywords}
	
	
	
\section{Introduction}{\label{intro}}

The accretion of gas onto black holes (BHs) is believed to be one of the primary sources of power for a wide range of active phenomena in our universe, such as X-ray binaries (XRBs), gamma-ray bursts and active galactic nuclei (AGNs) \cite[\text{e.g.,}][]{Lamb1973,Treves1988,Esin1997,Fryer1999,Davis2006,Wilkinson2009,Yuan2010,Veledina2013,Chatterjee2020}. In terms of their temperature, the accreting gas can be classified into two very distinct categories, namely cold and hot accretion flows. The cold accretion flow, commonly explained using either standard thin disc model \cite[]{Shakura1973} and/or the slim disc model \cite[]{Abramowicz1988}, are radiatively efficient, and remain optically thick. These models are characterized by high mass accretion rate usually exceeds the Eddington limit. Indeed, the cold accretion models with temperatures in the range of $10^4 - 10^7 ~{\rm K}$ successfully explain the spectrum of luminous AGNs \cite[]{Liu2012,Netzer2014}, black hole X-ray binaries (BX-XRBs) in the high-soft state \cite[]{Meyer2000,Dexter2012}, narrow-line Seyfert galaxies \cite[]{Mineshige2000,Wang2003,Haba2008}, and ultra luminous X-ray sources \cite[]{Watarai2001,Chen2004,Godet2012,Soria2015}.

On the contrary, in a hot accretion model with a low mass accretion rate, only a small fraction of the energy generated by turbulence is radiated away and most of the thermal energy is stored in the accretion flow, which is then advected into the BH. As a result, the temperature of the gas becomes extremely high although its density and scale height remain smaller in comparison with the well-known standard thin disc \cite[]{Shakura1973}. Hot accretion flows (HAFs), the subject of this study, have a drastically reduced radiative efficiency, leading to this model being referred to as a radiatively inefficient accretion flow (RIAF) \cite[]{Ichimaru1977,Narayan1994,Yuan2014}.

It is noteworthy that HAF models successfully explain the observational features of various BH systems including the supermassive BH in our Galactic center (Sgr A*) \cite[]{Manmoto1997,Yuan2002,Yuan2014}, M87 \cite[]{Reynolds1996,Park2019}, and the other low-luminosity AGNs (LLAGNs) \cite[]{Lasota1996,Nemmen2006,Nemmen2014,Younes2019}, and also BH-XRBs in the hard/quiescence states \cite[]{Esin1997,Hameury1997,Yuan2005,Liu2011}.

One of the most important findings of the numerical simulations is the existence of outflows in HAFs \cite[\text{e.g.,}][]{Ohsuga2009,Yuan2012a,Yuan2012b,Yuan2015,Bu2016a,Bu2016b,Mosallanezhad2022} that have been confirmed by observation of LLAGNs and XRBs \cite[\text{e.g.,}][]{Wang2013,Cheung2016,Homan2016,Ma2019,Park2019}. In the presence of outflows, mass, angular momentum, and energy are removed from the disc, which can have a profound effect on the dynamics and structure of the flow \cite[]{Yuan2018,Bu2019}. Therefore, the modelling of HAFs is able to make the properties of winds/outflows easier. For instance, a recent study of \cite{Yang2021} indicates that a larger BH spin and stronger magnetic fields lead to stronger winds/outflows from the disk.

Taking into account the temperature and density profiles of the HAFs with very low accretion rates, it appears that the collisional mean free paths of the charged particles are much larger than the typical length-scale of accretion flows, \text{i.e.,} $r_{\mathrm{g}} = GM_{\rm BH}/c^2$, where $ r_{\mathrm{g}} $ is the gravitational radius, and $ G $, $ M_{\rm BH} $, and $ c $ are the gravitational constant, the BH mass, and speed of light, respectively \cite[]{Mahadevan1997,Tanaka2006,Johnson2007}. The plasma in HAFs is therefore expected to be collisionless with thermal conduction playing a significant role.

The effect of thermal conduction on the physical properties of HAFs has been explored in several studies based on self-similar assumptions \cite[][]{Tanaka2006,Shadmehri2008,Faghei2012mnras,Khajenabi2013,Ghoreyshi2020,Mosallanezhad2021}. In an early attempt, \cite{Tanaka2006} reported that thermal conduction in HAFs possibly helps the gas to be launched from the disc as outflows. In addition, the effect of thermal conductivity on the energy flux of outflows as well as the size of the outflowing region appears to be significant \cite[]{Khajenabi2013}. Meanwhile, numerical simulations of HAFs indicate that the energy flux carried by the outflows in the presence of thermal conduction can be increased by a factor of $\sim 10$ \cite[]{Bu2011,Bu2016}. Further, \cite{Narayan1995a,Narayan1995b} suggested that the positive Bernoulli parameter is required for outflows to occur in an accretion disc. As a result, the gas becomes gravitationally unbound and escape from the gravitational potential of the central BH. It is noteworthy that a positive Bernoulli parameter results from the self-similar framework as well \cite[]{Nakamura1998,Yuan1999,Abramowicz2000, Yuan2015}.

Although the self-similar solutions provide the physical insights of the accretion flow, they fail to decipher the global behaviour of the accretion flow, especially at the inner and outer disc boundaries \cite[]{Narayan1997,Chen1997}. Because of this, several authors investigated the global solutions to HAFs around black holes in a self-consistent manner \cite[]{Abramowicz1996,Narayan1997,Chen1997,Nakamura1997,Popham1998,Lu1999,Becker2003,Chakrabarti-Das2004,Das-2007,Yuan2008,Das-etal2009,Narayan2011,Kumar2018,Kumar2021,Das-etal2022,Mitra-etal2022}. A pioneering and fascinating study of the global structure and dynamics considering single temperature HAFs was carried out by \cite{Narayan1997}. Upon comparing the global and self-similar solutions, they showed that the self-similar solutions satisfactorily mimic the regions avoiding the inner and the outer boundaries of the disc. Due to this, the spectra derived by using the self-similar solutions require modifications, as the swarm of high energy photons originated from the regions near the inner boundary are not accounted appropriately. In addition, \cite{Narayan1997} reported that the global solution leads to a negative Bernoulli parameter in the outer regions of the disc \cite[see also][]{Yuan1999,Kumar2018}. Further, \cite{Yuan1999} showed that the outer boundary conditions may significantly affect the value of Bernoulli parameter and its sign, as well. In reality, the Bernoulli parameter depends not only on the outer boundary conditions, but also on factors, such as the viscosity parameter, the adiabatic index, and the advection parameter \cite[]{Narayan1997,Popham1998,Narayan2011,Kumar2018}.

Over the course of accretion, the infall velocity approaches the speed of light when the accreting matter enters the BH \cite[]{Weinberg1972}, while it becomes negligible at large distances away from the BH horizon \cite[and references therein]{Frank-etal2002,Das-2007}. As a result, the accreting gas experiences a subsonic to supersonic transition at a point called the critical point \cite[]{Liang1980,Abramowicz1981}. The critical point depends on the value of the viscosity parameter and the outer boundary conditions \cite[]{Chakrabarti1996,Narayan1997,Yuan1999,Chakrabarti-Das2004,Yuan2008}. Moreover, if there are multiple critical points in accretion flows, the flows may undergo shocks \cite[]{Fukue1987}. The shocked disc may satisfy the observational criteria for the formation of the observed outflows \cite[]{Das-etal2001,Becker2008,Das-Chattopadhyay2008,Das-etal2009,Aktar-etal2015,Aktar-etal2017,Aktar-etal2018}. {In particular,}  \cite{Das-etal2009} demonstrated that the existence of shocks relies on the level of viscous dissipation. However, \cite{Narayan1997} examined similar transonic solutions for a wide range of the viscosity parameter, but did not report any shock \cite[see also][]{Chen1997,Nakamura1997,Lu1999,Yuan2008}, possibly due to the choice of selective boundary conditions.

Considering all these, in this work, we intend to examine the three primary objectives concerning the HAFs. Firstly, we aim to investigate the global transonic solutions of HAFs that include thermal conduction. This is particularly relevant for systems with an extremely low mass accretion rate, such as Sgr A* and the M87 galaxy, where the accretion flows are weakly collisional. In such systems, the electron collisional mean free path can be comparable to the typical size of the system, resulting in a significant influence of thermal conduction on the dynamics of the accretion flow and energy transport from the inner to outer regions \cite[]{Johnson2007,Quataert2008}. Our next objective is to determine the range of the thermal conduction parameter within which global solutions are viable for the given set of physical input parameters. This is an essential step in our study, as it allows us to identify the critical threshold for thermal conduction for which the global solutions cease to exist. Thirdly, we compare the results of global transonic solutions with the self-similar solutions of HAFs in the presence of thermal conductivity. This analysis provides a more detailed understanding of the impact of thermal conduction on the dynamics of HAFs.

The remainder of the manuscript is organized as follows. In Section \ref{sec:numerical_method}, the basic equations, physical assumptions, and the boundary conditions are introduced. The numerical results are presented in detail in Section \ref{sec:results}. Finally, in Section \ref{sec:summary_discussion}, we provide the discussion and summary of the present work.

\section{Hot Accretion Flows with Saturated Thermal Conduction} \label{sec:numerical_method}

We begin with a low angular momentum, steady, axisymmetric, viscous, advective accretion flow around a non-rotating black hole. Moreover, we assume that the mass accretion takes place at very low rate representing the radiatively inefficient hot accretion flow (HAF). In the subsequent sections, we study the properties of the HAF in the presence of thermal conduction.

\subsection{Dynamical equations} \label{subsec:dynamical_equations}

In order to deal with the HAF, we adopt a cylindrical coordinate system $ (r, \phi, z) $. We employ the same set of height-integrated governing equations as delineated in \cite{Narayan1997} except the energy equation, where we include the effect of thermal conduction. In addition, we consider the hydrostatic equilibrium in the vertical direction and hence, the flow variables are vertically averaged. Accordingly, in this formulation, the flow variables are expressed as functions of the cylindrical radius $r$ only. Under these assumptions, the governing equations are given by,
\begin{equation} \label{eq:continuity}
	\dot{M} = - 4 \pi r H \rho v,	
\end{equation}

\begin{equation} \label{eq:mom1}
	v \frac{d v}{d r} =  (\Omega^{2} - \Omega_\mathrm{_K}^{2} ) r - \frac{1}{\rho} \frac{d (\rho C_{\rm s}^{2})}{d r},	
\end{equation}

\begin{equation} \label{eq:mom3}
	\rho r H v \frac{d(\Omega r^{2})}{d r} =  \frac{d}{d r} \left( \nu \rho H r^{3} \frac{d \Omega}{d r} \right),	
\end{equation}

\begin{equation} \label{eq:energy}
	 \frac{\rho v}{(\gamma -1)} \frac{d C_{\rm s}^{2}}{d r} - C_{\rm s}^{2} v \frac{d \rho}{d r} =
	 f \nu \rho r^{2} \left( \frac{d \Omega}{d r} \right)^{2} - \frac{1}{r} \frac{d ( r F_{\rm s})}{d r}.	
\end{equation}
In the above equations, $ \rho $ and $ \Omega $ are the mass density and the angular velocity of the gas, respectively. The radial velocity of the flow $ v $ is assumed to be negative for an inward flow of gas. Here, $ H \equiv C_{\rm s} / \Omega_{_{\rm K}} $ is the vertical half-thickness of the flow, where $ C_{\rm s} $ is the isothermal sound speed and $ \Omega_{_{\rm K}} $ is the Keplerian angular velocity. Adopting pseudo-Newtonian potential $ \Psi=-GM_{\rm BH}/(r-r_{\rm s}) $ \cite[]{Paczynsky1980}, the Keplerian angular velocity is given by,
\begin{equation} \label{eq:omgK}
	\Omega_{_{\rm K}}^2 = \frac{GM_{\rm BH}}{r(r-r_{\rm s})^2},
\end{equation}
where $ r_{\rm s} = 2GM_{\rm BH}/c^2 $ is the Schwarzschild radius for a BH with mass $M_{\rm BH}$. The last term on the right hand side of equation (\ref{eq:mom1}) is the acceleration due to the pressure gradient. Here, the pressure is defined by the isothermal sound speed $ C_{\rm s} $ and the density $ \rho $ as $ p=\rho C_{\rm s}^2 $. A Shakura-Sunyaev prescription \cite[]{Shakura1973} is adopted for the kinematic coefficient of viscosity ($ \nu $) as,
\begin{equation} \label{eq:nu}
	\nu = \alpha C_{\rm s} H,
\end{equation}
where $ \alpha $ is the viscosity parameter. We assume that the viscosity parameter is a constant, and is independent of $ r $. By substituting equation (\ref{eq:nu}) into equation (\ref{eq:mom3}), and using equation (\ref{eq:continuity}), we have,
\begin{equation}
	\frac{d}{d r}\left( \rho H v r^{3} \Omega \right) =  \frac{d}{d r} \left( \frac{\alpha C_{\rm s}^{2} \rho H r^{3}}{\Omega_{_{\rm K}}} \frac{d \Omega}{d r} \right),
\end{equation}
which on integration gives
\begin{equation} \label{eq:domg}
	\frac{d \Omega}{d r} = \frac{v \Omega_{_{\rm K}} ( \ell - \ell_0 )}{\alpha r^{2} C_{\rm s}^{2}},
\end{equation}
where $ \ell = \Omega r^{2} $ is the angular momentum per unit mass (hereafter specific angular momentum) for the accreting gas at radius $ r $. The integration constant $ \ell_0 $ represents the specific angular momentum eventually swallowed by the black hole.
In energy equation (\ref{eq:energy}), $ \gamma $ is the ratio of specific heats of the gas. The advection parameter $ f~(=1-Q_{\rm rad}/Q_{\rm vis}) $ is assumed to be a constant which lies in the range $0 \le f \le1$. Here, $Q_{\rm vis}$ and $Q_{\rm rad}$ denote viscous heating and radiative cooling rates. Since the collisional mean free paths of the charged particles in HAFs are much larger than the typical length-scale of the accretion flows, one may no longer apply the classical theory for thermal conduction. Under these conditions, the heat flux is described as the saturated form of conduction. The last term on right hand side of equation (\ref{eq:energy}) represents the transfer of energy due to the saturated
thermal conduction. Following \cite{Cowie1977}, the saturated conduction flux $ F_{\rm s} $ is obtained as,
\begin{equation} \label{eq:Fs}
F_{\rm s} = 5 \Phi_{\rm s} \rho C_{\rm s}^{3},
\end{equation}
where $ \Phi_{\rm s} $ is the dimensionless saturation constant with $0 \le \Phi_{\rm s} < 1$. It is noteworthy that the self-similar solutions describing the accretion flow tend to become non-rotating ($\Omega \rightarrow 0$) when the saturation constant ($ \Phi_{\rm s}$) reaches its limiting value ($\Phi_{\rm sc}$) \cite[]{Shadmehri2008,Ghasemnezhad2012,Faghei2012,Ghoreyshi2020}. Accordingly, the physically acceptable accretion solutions around a black hole are given by the remaining allowed range of the saturation constant, $0 \le \Phi_{\rm s} \le \Phi_{\rm sc}$.

\subsection{Critical point and boundary conditions} \label{subsec:BC}

Using equations (\ref{eq:continuity}), (\ref{eq:energy}), (\ref{eq:domg}), and (\ref{eq:Fs}), we get the radial gradient of the sound speed as,

\begin{multline} \label{eq:dcs}
\left(  \frac{\gamma + 1}{\gamma - 1} + 10\, \Phi_{\rm s} \frac{C_{\rm s}}{v} \right) \frac{d \ln C_{\rm s}}{d r} = - \left( 1 - 5 \Phi_{\rm s} \frac{C_{\rm s}}{v} \right) \frac{d \ln |v|}{dr} \\
~~~~~~~~~~~~~~~~~~~~~~~~~~~~~~~+ \left( 1 - 5 \Phi_{\rm s} \frac{C_{\rm s}}{v} \right) \frac{d \ln \Omega_{\rm _K}}{d r} - \frac{1}{r} + \frac{f v \Omega_K}{\alpha r^2 C_{\rm s}^{4}}\big( \ell - \ell_0 \big)^2.
\end{multline}

We next use equations (\ref{eq:continuity}) and (\ref{eq:dcs}) to eliminate $ d \rho / d r $ and $ d C_{\rm s} / d r $ in equation (\ref{eq:mom1}) and thereby express the differential dynamical equation as,

\begin{multline} \label{eq:dv}
\left[ \frac{ 2\gamma + 5\, \Phi_{\rm s} ( \gamma - 1 ) C_{\rm s} / v }{( \gamma + 1 ) + 10\, \Phi_{\rm s} ( \gamma - 1 ) C_{\rm s} / v } -\frac{v^{2}}{C_{\rm s}^2}  \right]  \frac{d \ln |v|}{d r} \\
= \frac{r \left( \Omega_{\rm _K}^{2} - \Omega^{2} \right)}{C_{\rm s}^{2}} - \left[ \frac{ 2\gamma + 10\, \Phi_{\rm s} ( \gamma - 1 ) C_{\rm s} / v }{( \gamma + 1 ) + 10\, \Phi_{\rm s} ( \gamma - 1 ) C_{\rm s} / v } \right] \frac{1}{r} \\
+ \left[ \frac{ 2\gamma + 5\, \Phi_{\rm s} ( \gamma - 1 ) C_{\rm s} / v }{( \gamma + 1 ) + 10\, \Phi_{\rm s} ( \gamma - 1 ) C_{\rm s} / v} \right] \frac{d \ln \Omega_{\rm _K}}{d r}+ \frac{ f \Omega_{\rm _K} v }{\alpha r^{2} C_{\rm s}^{4}}\\
\times\left[ \frac{\gamma - 1}{ ( \gamma + 1 ) + 10\, \Phi_{\rm s} ( \gamma - 1 ) C_{\rm s} / v } \right] \left( \ell - \ell_0 \right)^{2}.
\end{multline}

We numerically solve the differential equations (\ref{eq:domg}), (\ref{eq:dcs}), and (\ref{eq:dv}) to obtain the radial profile of $v$, $C_s$ and $\ell$. In doing so, one requires to supply the boundary conditions. As stated in the introduction, the inflowing gas starts its journey from the outer edge of the disc with negligible radial velocity ($ |v| \ll c $, \text{i.e.,} subsonic). However, the matter flows into the BH with supersonic velocity ($ |v| \sim c $) to satisfy the inner boundary conditions imposed by the event horizon. Therefore, the flow must change its sonic state at the critical point ($ r_c $) to become transonic at least once, if not multiple times. At the critical point, the radial velocity gradient takes the form $ dv/dr |_{\rm c} =0/0$ as both numerator $ {\cal N}_{\rm c} $ and denominator ${\cal D}_{\rm c}$ simultaneously vanish at $r_c$, and we have the critical point conditions ${\cal N}_{\rm c}={\cal D}_{\rm c}=0$, which are explicitly yielded as,
\begin{equation}\label{eq:D}
\mathcal{D}_{\rm c} \equiv \frac{ 2\gamma + 5\, \Phi_{\rm s} ( \gamma - 1 ) C_{\rm sc} / v_{\rm c} }{( \gamma + 1 ) + 10\, \Phi_{\rm s} ( \gamma - 1 ) C_{\rm sc} / v_{\rm c} } -\frac{v_{\rm c}^{2}}{C_{\rm sc}^2} = 0,
\end{equation}
\begin{multline}\label{eq:N}
\mathcal{N}_{\rm c} \equiv \frac{r_{\rm c} \left( \Omega_{\rm _K}^{2} - \Omega_{\rm c}^{2} \right)}{C_{\rm sc}^{2}} - \left[ \frac{ 2\gamma + 10\, \Phi_{\rm s} ( \gamma - 1 ) C_{\rm sc} / v_{\rm c} }{( \gamma + 1 ) + 10\, \Phi_{\rm s} ( \gamma - 1 ) C_{\rm sc} / v_{\rm c} } \right] \frac{1}{r_{\rm c}} \\
+ \left[ \frac{ 2\gamma + 5\, \Phi_{\rm s} ( \gamma - 1 ) C_{\rm sc} / v_{\rm c} }{( \gamma + 1 ) + 10\, \Phi_{\rm s} ( \gamma - 1 ) C_{\rm sc} / v_{\rm c}} \right] \frac{d \ln \Omega_{\rm _K}}{d r} + \frac{ f \Omega_{\rm _K} v_{\rm c} }{\alpha r_{\rm c}^{2} C_{\rm sc}^{4}} \\
\times\left[ \frac{\gamma - 1}{ ( \gamma + 1 ) + 10\, \Phi_{\rm s} ( \gamma - 1 ) C_{\rm sc} / v_{\rm c} } \right] \left( \ell_{\rm c} - \ell_0 \right)^{2} = 0.
\end{multline}
where $ v_{\rm c} $, $ \Omega_{\rm c} $, $ C_{\rm sc} $, and $ \ell_{\rm c} $ denote the radial, and the angular velocities, sound speed and the angular momentum at the critical point ($r_c$), respectively. Since the flow remains smooth along the streamline, $dv/dr$ must be real and finite all throughout. Hence, we calculate $dv/dr|_{\rm c}$ by applying the l$'$H\^{o}pital's rule, leading to
\begin{equation} \label{eq:dvcp}
\bigg(\frac{d v}{d r} \bigg)_{\rm c} = \bigg( \frac{ d \mathcal{N} / d r }{ d \mathcal{D} / d r } \bigg)_{r=r_{\rm c}}.
\end{equation}
In general, $ dv/dr |_{\rm c} $ possesses two distinct values at $ r_{\rm c} $. When both values of $ dv/dr |_{\rm c} $ are real and of opposite sign, $i.e.$, $ dv/dr |_{\rm c}  < 0 $, and $ dv/dr |_{\rm c} > 0 $, we obtain saddle type critical points \cite[and referenceses therein]{Das-2007,Das-etal2009,Mitra-etal2022}. Note that saddle type critical points are of special interest as the global transonic accretion flow has to pass through it \cite[]{Chakrabarti-Das2004}.

Another boundary condition implies the vanishing of the viscous shear stress at the horizon \cite[]{Becker2003, Das-etal2009}. Hence, considering $ d \Omega / d r = 0 $, we obtain,
\begin{equation}
	\lim_{r \to r_{\rm s}} \Omega(r) \equiv \Omega_0 = \frac{\ell_0}{r_{\rm s}^2}.
\end{equation}

Applying the aforementioned conditions and adopting the methodology outlined in the following subsection, we obtain the comprehensive global transonic accretion solutions around black holes in the presence of thermal conduction. With careful adherence to this approach, we can accurately model the behaviour and characteristics of the accretion flow that provides the precious insights into the underlying physical processes under considerations.

\subsection{Globally conserved energy equation} \label{subsec:energy constant}

In order to obtain the energy transport rate per unit mass of a viscous advective flow in the presence of thermal conduction, we rewrite equation (\ref{eq:energy}) as
\begin{multline} \label{eq:conservedE}
    \mathcal{E}=\frac{v^2}{2}- \frac{\ell^2}{2r^2} + \Psi + h +\frac{\ell \ell_0}{r^{2}}+\frac{ 5 \Phi_{\rm s} C_{\rm s}^{3} }{v}  \\
    -\int{\bigg(\frac{5 \Phi_{\rm s} C_{\rm s}^{3}}{v H} \frac{dH}{dr}\bigg) dr},
\end{multline}
where $ h ~[=\gamma p/\rho(\gamma-1) ]$ is the specific enthalpy.
We note that the saturated conduction flux decreases in regions of the accreting flow where the electrons become relativistic \cite[]{Tanaka2006}. Therefore, the effect of thermal conduction can be negligible in the inner regions ($ r_{\rm in} $) of the disc. Accordingly, we set $ \Phi_{\rm s} \sim 0 $ at $ r_{\rm in} $, and redefine the energy transport rate per unit mass ($\varepsilon_0$) at $ r_{\rm in} $ as
\begin{equation} \label{eq:Econs}
    \varepsilon_0 = \frac{v^2}{2}- \frac{\ell^2}{2r^2} + \Psi + h +\frac{\ell \ell_0}{r^{2}}.
\end{equation}
We fix the energy $\varepsilon_0$ at $r_{\rm in}$ and obtain the global solutions following the methodology as delineated in Appendix \ref{Method}. It is important to note that $\varepsilon_0$ is conserved for a viscous, advective accretion flow. Moreover, we express the Bernoulli parameter ($Be$) \cite[]{Nakamura1997} that contains the local information of radial motion, azimuthal motion, gravity and thermodynamic terms, and is given by,
\begin{equation} \label{eq:Be}
    Be = \frac{v^2}{2} + \frac{\ell^2}{2r^2} + \Psi + h.
\end{equation}
Needless to mention that in absence of any viscosity, $\ell = \ell_0$ and hence, we have $\varepsilon_0 = Be$.

\section{Numerical Results} \label{sec:results}

In obtaining the accretion solutions, we employ a unit system with $ GM_{\rm BH}=c=1 $. This allows us to simplify our calculations and obtain results that are seamlessly compared with previous studies. The structure of HAFs is influenced by a variety of factors, including the energy transport rate $\varepsilon_0$ at $r_{\rm in}$, angular momentum transport rate $\ell_0$ at the horizon, the viscosity parameter $\alpha$, the saturation constant $\Phi_{\rm s}$, and the ratio of specific heats, $\gamma$. In this study, we choose  $\gamma=1.5$ unless stated otherwise, and set $f=1$ for the purpose of representation. To find the critical point location, we utilize the iteration methodology as described in \cite{Becker2003,Das-etal2009,Kumar2018} (see appendix \ref{Method} for more details). We then solve the coupled differential equations (\ref{eq:domg}), (\ref{eq:dcs}), and (\ref{eq:dv}) simultaneously for a given set of input parameters ($\varepsilon_0,\ell_0,\alpha,\Phi_{\rm s}$) to obtain the global solution for HAFs in the presence of thermal conduction. This approach allows us to accurately model the behaviour of HAFs and gain insights into the complex processes that govern their structure.

\begin{figure}
	\includegraphics[width=\columnwidth]{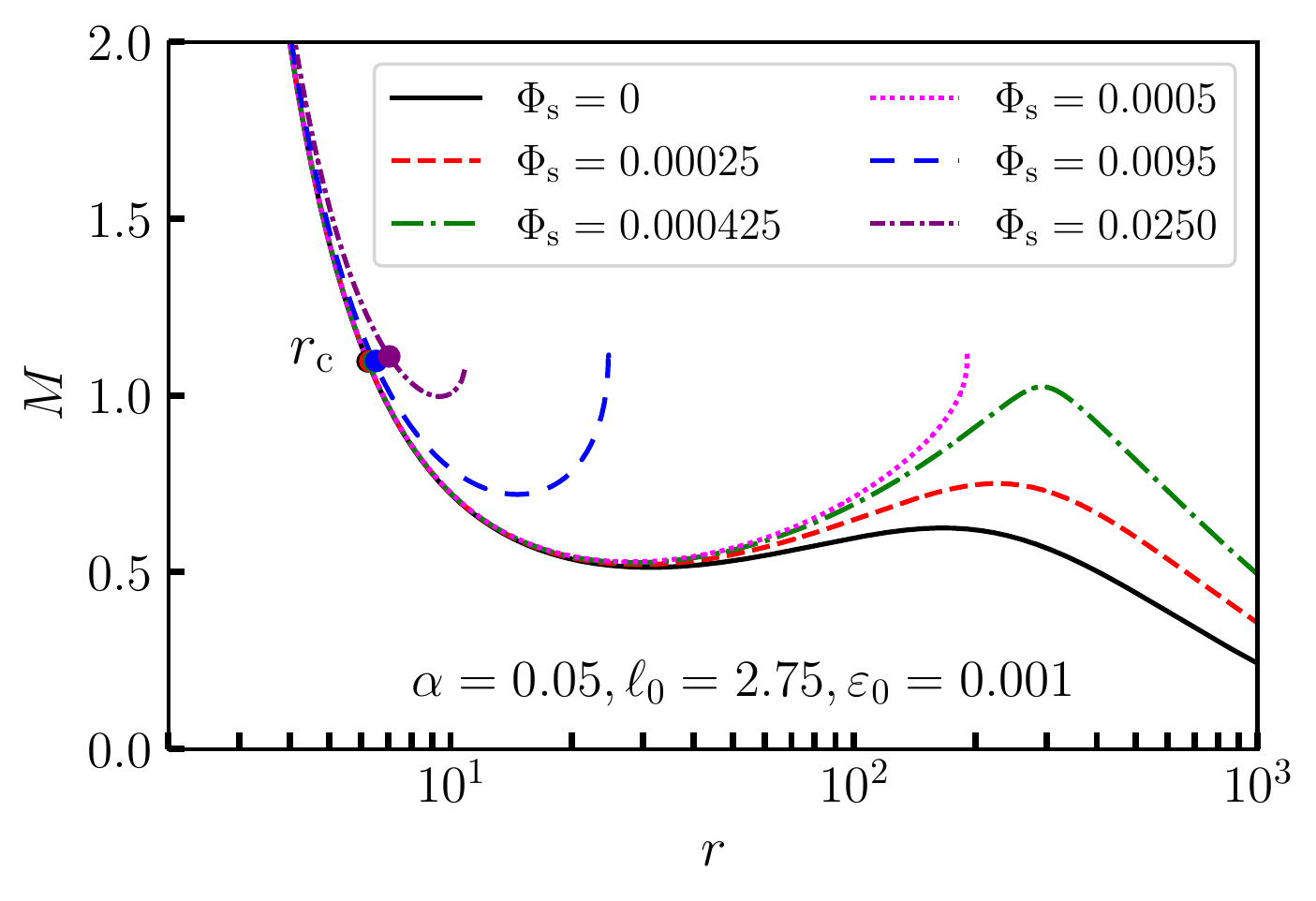}
	\caption{Variation of Mach number ($M=|v|/C_{\rm s}$) as a function of radial coordinate ($r$) for different $ \Phi_{\rm s} $ values starting from $0$ to $0.0250$ which are marked. Here, the input parameters are chosen as $(\varepsilon_0,\ell_0,\alpha)=(0.001,2.75,0.05)$. The filled circles represent the critical points. See text for the details.
	}
\label{fig:Mach_conduction1}
\end{figure}

\subsection{Global transonic solutions} \label{subsec:global_solutions}

We choose a set of input parameters, $ (\varepsilon_0,\ell_0,\alpha) = (0.001,2.75,0.05) $, and integrate the flow equations (\ref{eq:domg}, \ref{eq:dcs}, \ref{eq:dv}) towards the outer edge ($r_{\rm edge}$) of the disc starting from $ r_{\rm in} = 2.001$ considering $\Phi_{\rm s} = 0 $. The obtained results are depicted in Fig. \ref{fig:Mach_conduction1}, where the solid (black) curve smoothly connects the horizon with $r_{\rm edge}=1000$ via a critical point at $r_{\rm c}=6.233$. Solutions of this kind where a sub-sonic flow ($v \ll c$) from a large distance smoothly crosses the BH horizon supersonically are called global accretion solutions. Next, we increase the saturation constant to $ \Phi_{\rm s} = 0.00025$, and notice that the obtained global solution (dashed curve in red) deviates from the global solution with $\Phi_{\rm s} = 0 $. It is interesting to note that the global accretion solutions obtained for different $ \Phi_{\rm s}$ remain quite insensitive, particularly in the inner regions of the disk. However, the effect of thermal conduction on the accretion solutions is prominently visible in the region far from the black hole horizon. We keep increasing the saturation constant to a critical value $ \Phi_{\rm s} = 0.000425 $ (dot-dashed curve in green), beyond that the flow fails to connect the outer edge as the solution becomes closed \cite[]{Sarkar-Das2018}, shown using a dotted (magenta) curve. If we keep increasing $ \Phi_{\rm s} $, we continue to obtain closed solutions depicted in long-dashed (blue) and dot-dashed (purple) curves. Note that these solutions are apparently unphysical unless they join via shock with other solutions passing through another critical point usually located far away from the horizon \cite[and references therein]{Fukue1987,Chakrabarti-1989,Chakrabarti1996,Das-etal2001,Chakrabarti-Das2004,Das-2007,Das-etal2009,Das-etal2022}. Finally, we find an upper limit of the saturation constant, $\Phi_s = 0.025$, above which accretion solutions cease. We observe that the critical point shifts outwards when increasing the saturation constant (see \S~\ref{sonic_props}). This finding contradicts the previously reported results \cite[]{Faghei2012mnras} and hence, we intend to analyse this in detail in the following subsection \S~\ref{sonic_props}.

\begin{figure}
	\centering
	\includegraphics[width=\columnwidth]{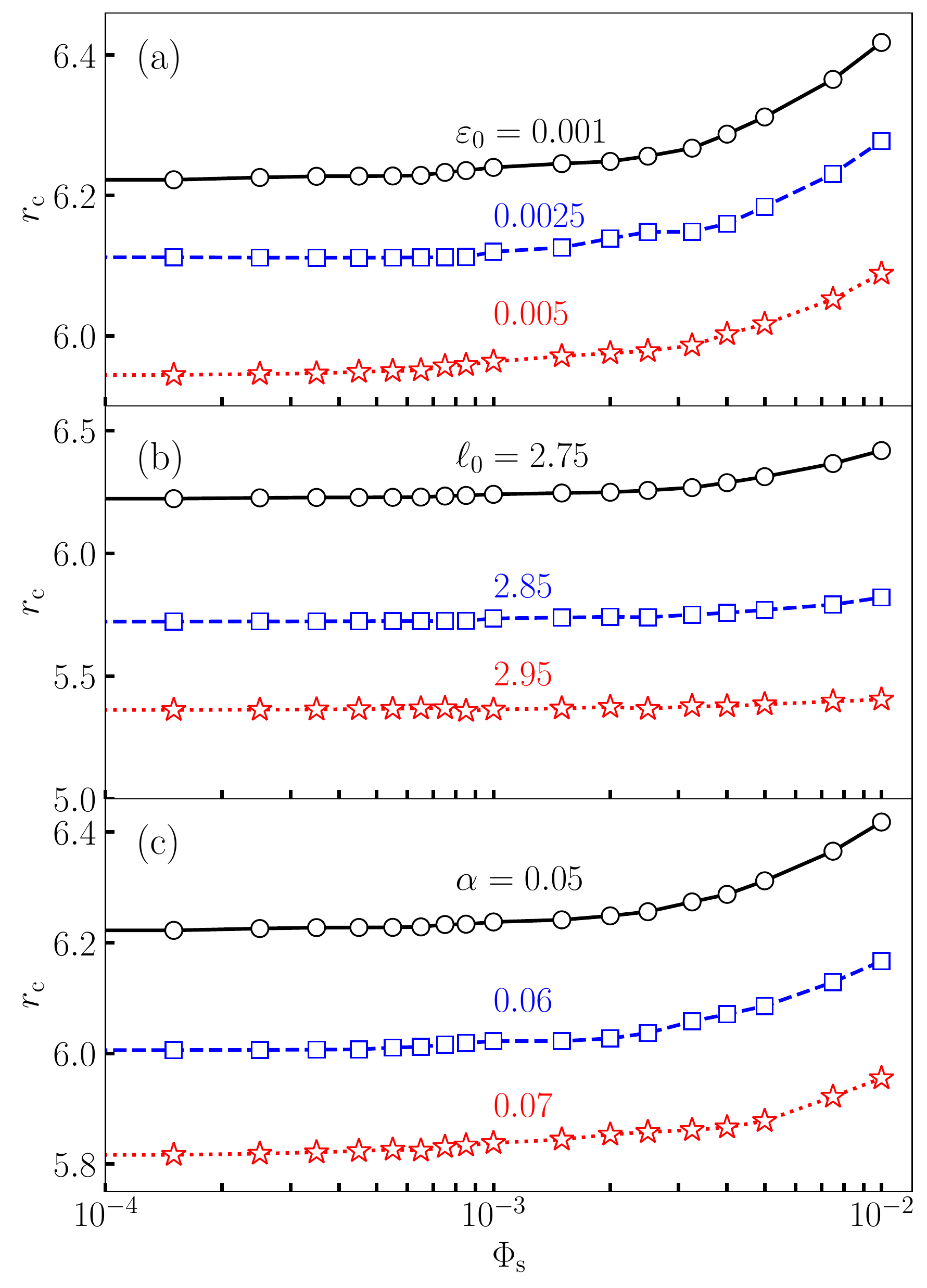}
	\caption{Variation of the critical point location $ (r_{\rm c}) $ as a function of $ \Phi_{\rm s} $. In panel (a), we fix $ (\alpha, \ell_0)=(0.05, 2.75) $, and show $ r_{\rm c} $ for $\varepsilon_0 = 0.001, 0.0025$, and $0.005 $. In panel (b), we choose $(\varepsilon_0, \alpha)=(0.001,0.05) $, and obtain results for different angular momentum at the horizon as $ \ell_0 = 2.75,2.85$, and $2.95 $. In panel (c), we set ($ \varepsilon_0, \ell_0 $) = $ (0.001,2.75) $, and vary the viscosity parameter as $ \alpha = 0.05, 0.06$, $0.07 $. In each panel, open circles, squares and asterisks represent the location of critical points $ r_{\rm c} $. See text for the details.
	}
	\label{fig:rc}
\end{figure}

\subsection{Dependency of critical point on input parameters}\label{sonic_props}

\begin{figure}
	\includegraphics[width=\columnwidth]{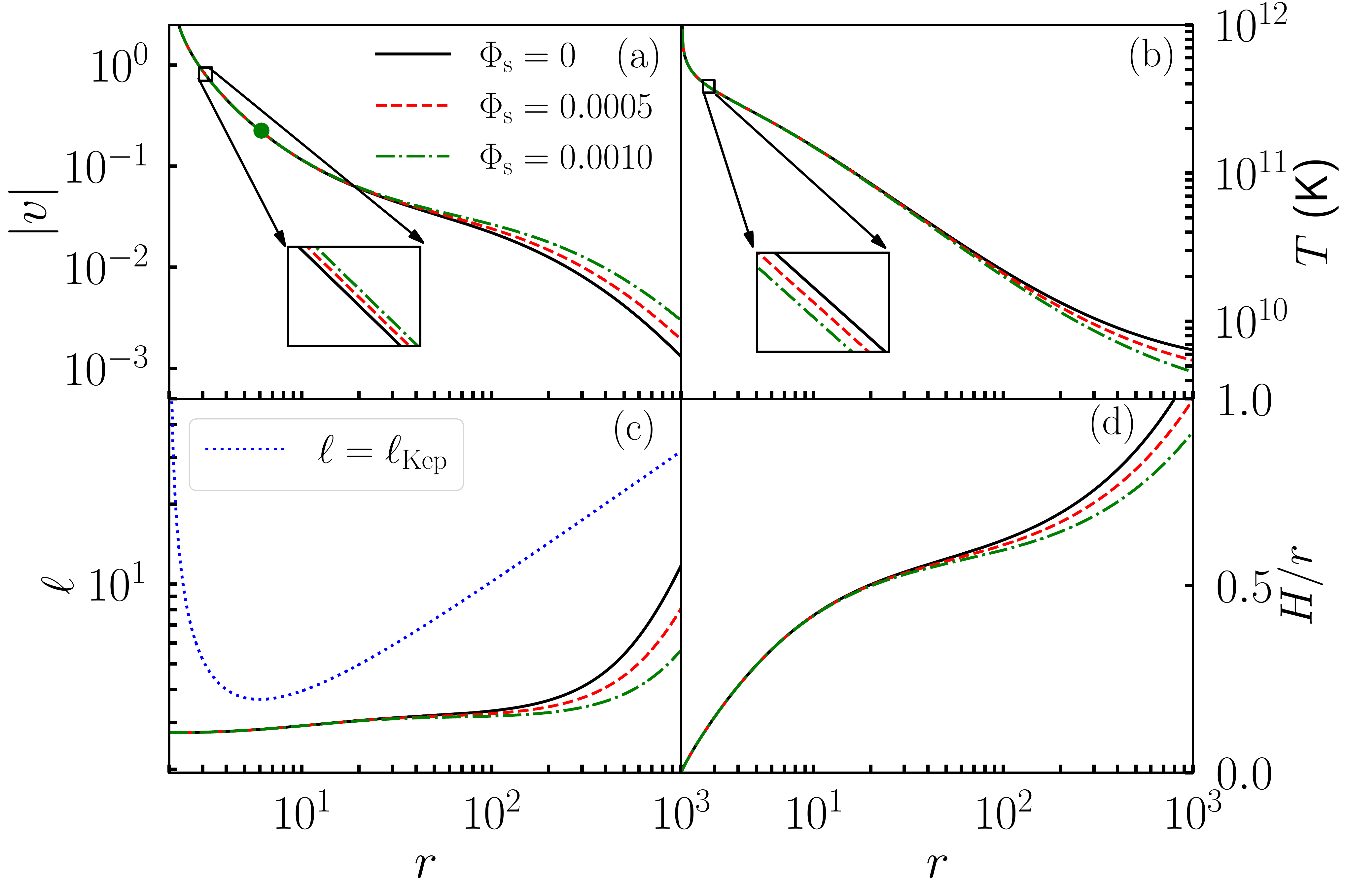}
	\caption{The profiles of velocity $ v $, temperature $ T $, angular momentum $ \ell $, and aspect ratio $ H/r $ are plotted as a function of radial distance $ r $ for different values of $ \Phi_{\rm s} = 0$ (solid), $0.0005$ (dashed), and $0.0010$ (dot-dashed), respectively. Here, the input parameters are chosen as $(\varepsilon_0,\ell_0,\alpha)=(0.0025,2.75,0.05) $. See text for the details.
	}
\label{fig:variables}
\end{figure}

In Fig. \ref{fig:rc}, we find a unique correspondence between the critical point location and the saturation constant ($ \Phi_{\rm s}$) for different combinations of $(\varepsilon_0,\ell_0,\alpha)$. In Fig. \ref{fig:rc}a, we choose $(\varepsilon_0,\ell_0,\alpha)=(0.001, 2.75,0.05)$, and start with $ \Phi_{\rm s} = 0 $. For this configuration, we obtain the critical point at $ r_{\rm c} = 6.233 $, and as $\Phi_{\rm s} $ is increased, the critical point shifts outwards. In reality, as $\Phi_s$ is increased, flow temperature at a given radial coordinate is decreased (see Fig. \ref{fig:variables}b for more details), and hence, $C_s$ is also decreased there. Further, since Mach number $M_{r_{\rm c}} ~[= (v/C_s)_{r_{\rm c}}]$ at $r_{\rm c}$ remains largely insensitive to $\Phi_s$, $r_{\rm c}$ shifts outward with the increase of $\Phi_s$ to restore $M_{r_{\rm c}}$. This result is shown using open circles joined using solid (black) lines. Next, we keep $(\ell_0, \alpha)$ fixed, and increase energy to $ \varepsilon_0 = 0.0025, 0.005$ that causes the critical point location to reduce (see open squares in blue and open asterisks in red). Indeed, as energy is increased, the temperature of the disc is also increased, which causes the critical points to move inwards to maintain the higher temperature. In Fig. \ref{fig:rc}b, we fix ($ \varepsilon_0, \alpha $) = ($ 0.001, 0.05 $) and vary angular momentum as $ \ell_0 = 2.75, 2.85 $, and $2.95$, respectively. Finally, in Fig. \ref{fig:rc}c, we only vary the viscosity parameter as $ \alpha = 0.05, 0.06$, and $0.07$ keeping other parameters fixed. When $ \ell_0 $ or $ \alpha $ is increased, the frictional force increases within the flow that eventually yielding enhanced viscous heating. Hence, the critical points move inwards with the increase of $ \ell_0$ or $\alpha $, although we observe an anti-correlation between $ \Phi_{\rm s} $ and any one of the global input parameters, namely $\varepsilon_0, \ell_0$, and $\alpha $, over the variation of $ r_{\rm c} $. However, the overall variation of the critical point location ($r_{\rm c}$) with $ \varepsilon_0, \ell_0$, and $\alpha$ appears to remain similar as was reported earlier \cite[]{Chakrabarti-Das2004,Das-etal2009,Das-etal2022,Mitra-etal2022}.

\subsection{The effect of thermal conduction on flow variables}

In Fig. \ref{fig:variables}, we depict the behaviour of flow variables corresponding to global transonic solution in the presence of thermal conduction. Here, we set the input parameters as $(\varepsilon_0,\ell_0,\alpha)=(0.0025,2.75,0.05)$. The solutions are illustrated for different values of the saturation constant as $ \Phi_{\rm s} =0.0, 0.0005$, and $0.0010$, which are plotted using solid (black), dashed (red), and dot-dashed (green) curves, respectively. In panel (a), the sub-sonic accretion flow from $ r_{\rm edge}=1000 $ starts accreting with negligible velocity and gradually gains radial velocity as it proceeds towards the black hole. At $ r_{\rm c}$, flow becomes supersonic and ultimately crosses the horizon supersonically. Note that flow velocity exceeds the speed of light just outside the horizon. This happens due to the limitation of the pseudo-Newtonian potential which deviates to mimic the space-time geometry of the black hole there. For $ \Phi_{\rm s} = 0.0, 0.0005$, and $0.0010$, the critical points are obtained at $ r_{\rm c} = 6.0924$, $6.1113$, and $6.1219$, respectively. We find that radial velocity is increased marginally with $ \Phi_{\rm s}$ at the inner part of the disc shown at the inset, however, noticeably deviation is observed towards the outer part of the disc. In a convergent flow, the temperature ($T$) is increased with the decrease of $r$ mainly due to the geometrical compression. However, the presence of thermal conduction generally leads to the reduction of temperature, because the heat generated by the viscous dissipation is transferred away due to the thermal conduction. As expected, the reduction of temperature at the outer edge of the disc is observed (see panel (b) of Fig. \ref{fig:variables}), which are in agreement with the results of the numerical simulation \cite[]{Wu-etal2010}. In panel (c), we display the variation of the angular momentum $ \ell $ with $r$ corresponding to the solutions presented in panel (a). We find that the angular momentum transport is very inefficient particularly at the inner part of the disc, although the increase of $ \ell $ is seen at higher radial coordinates. Meanwhile, \cite{Faghei2012mnras} argued that for enhanced $\Phi_{\rm s}$, viscous turbulence is reduced that weakens the efficiency of angular momentum transport inside the disc. We further compare the flow angular momentum profile with the Keplerian angular momentum ($\ell_{\rm Kep} $) distribution (dotted curve in blue) and observe that $ \ell $ of HAFs remains sub-Keplerian all throughout. In panel (d), we demonstrate the relative thickness of the disc $ H/r $ at all radii. From the figure, it is clear that $ H/r \ll 1$ is generally maintained at the inner region, however, flow is intended to become quasi-spherical $H/r \sim 1$ towards the outer regions. Moreover, we find that the disc thickness is reduced at the outer regions as the influence of thermal conduction is increased. This is naturally expected, as the increased $ \Phi_{\rm s} $ generally reduces the disc temperature ($T$) at the outer part of the disc that eventually resulted the reduction of the disc height.

\subsection{Self-similar solutions} \label{subsec:selfsimilar}

In this paper, our main objective is to study the global transonic solutions of HAFs in presence of thermal conduction. In addition, we also intend to conduct a comparative analysis of HAFs by means of the self-similar solutions \cite[]{Narayan1994}. These analyses provide the valuable insights of the similarities and differences between the two solutions. Indeed, the self-similar solutions satisfactorily describe the structure of an accretion flow far from boundaries, and hence, such solutions are obtained for $ r \gg r_{\rm s} $ that reduces the pseudo-Newtonian potential in the Newtonian form. Following \cite{Narayan1994}, we choose the self-similar treatment in the following forms
\begin{equation}\label{eq:v}
	v(r) = - \alpha C_1 v_{\rm _K},
\end{equation}
\begin{equation}\label{eq:o}
	\Omega(r) = C_2 \Omega_{\rm _K},
\end{equation}
\begin{equation}\label{eq:cs}
	C_{\rm s}^2(r) = C_3 v_{\rm _K}^2,
\end{equation}
where $v_{\rm _K}~(=\sqrt{GM_{\rm BH}/r},~G=M_{\rm BH}=1)$ is the Keplerian velocity, and $C_1$, $C_2$ and $C_3$ are constants. By substituting the self-similar solutions (equations (\ref{eq:v}), (\ref{eq:o}), (\ref{eq:cs})) into the equations (\ref{eq:mom1})-(\ref{eq:energy}), we obtain a closed set of dimensionless equations that allow us to determine the constants $C_1$, $C_2$, and $C_3$. The closed set of dimensionless equations are given by,
\begin{equation}\label{eq:radial-momentum1}
	-\frac{1}{2} \alpha^2 {C_1}^2 = {C_2}^2 -1 +\frac{5}{2} C_3,
\end{equation}
\begin{equation}\label{eq:azimuthal-momentum1}
	C_1 = \frac{3}{2} C_3,
\end{equation}
\begin{equation}\label{eq:energy2}
	\Big[ \frac{1}{\gamma-1} - \frac{3}{2} \Big] C_1 = \frac{9}{4} f {C_2}^2 + 10 \frac{\Phi_{\rm s}} {\alpha} \sqrt{C_3}.
\end{equation}
After some algebraic manipulations, an equation for $C_1$ is obtained as
\begin{equation}\label{eq:algebraic-eq}
	\frac{9 f \alpha^2}{8} {C_1}^2 + \Big[ \frac{1}{\gamma-1} - \frac{3}{2} + \frac{15 f}{4} \Big] C_1 - \frac{10 \sqrt{6} \Phi_{\rm s}} {3 \alpha} \sqrt{C_1} - \frac{9 f}{4}=0.
\end{equation}
As reported in \cite{Tanaka2006}, that the solution of equation (\ref{eq:algebraic-eq}) yields the significant changes in the radial and rotational velocity profiles when thermal conduction is active inside the flow. In particular, they pointed out that in the presence of thermal conduction, the accreting flow  rotates with lower rate, while its inward motion becomes faster. Meanwhile, we mention in $\S$\ref{subsec:dynamical_equations} that the accreting flow reaches a non-rotating limit at a specific saturation constant $\Phi_{\rm sc}$. Accordingly, we calculate $\Phi_{\rm sc}$ using equations (\ref{eq:radial-momentum1})-(\ref{eq:energy2}) subject to the condition $C_2=0$. With this, we have
\begin{equation}\label{eq:phialpha}
	\Phi_{\rm sc}= \sqrt{-10+2\sqrt{18\alpha^2+25}}\frac{5-3\gamma}{40(\gamma-1)}.
\end{equation}
Equation (\ref{eq:phialpha}) clearly indicates that $\Phi_{\rm sc}$ strictly depends on both viscosity parameter $\alpha$ and ratio of specific heats $\gamma$, when self-similar solutions are adopted. What is more is that for $\Phi_{\rm s} > \Phi_{\rm sc}$, ${C_2}^2$ in equation (\ref{eq:energy2}) becomes negative resulting unphysical solutions as it leads to $\Omega^2 < 0$.

\begin{figure}
	\includegraphics[width=\columnwidth]{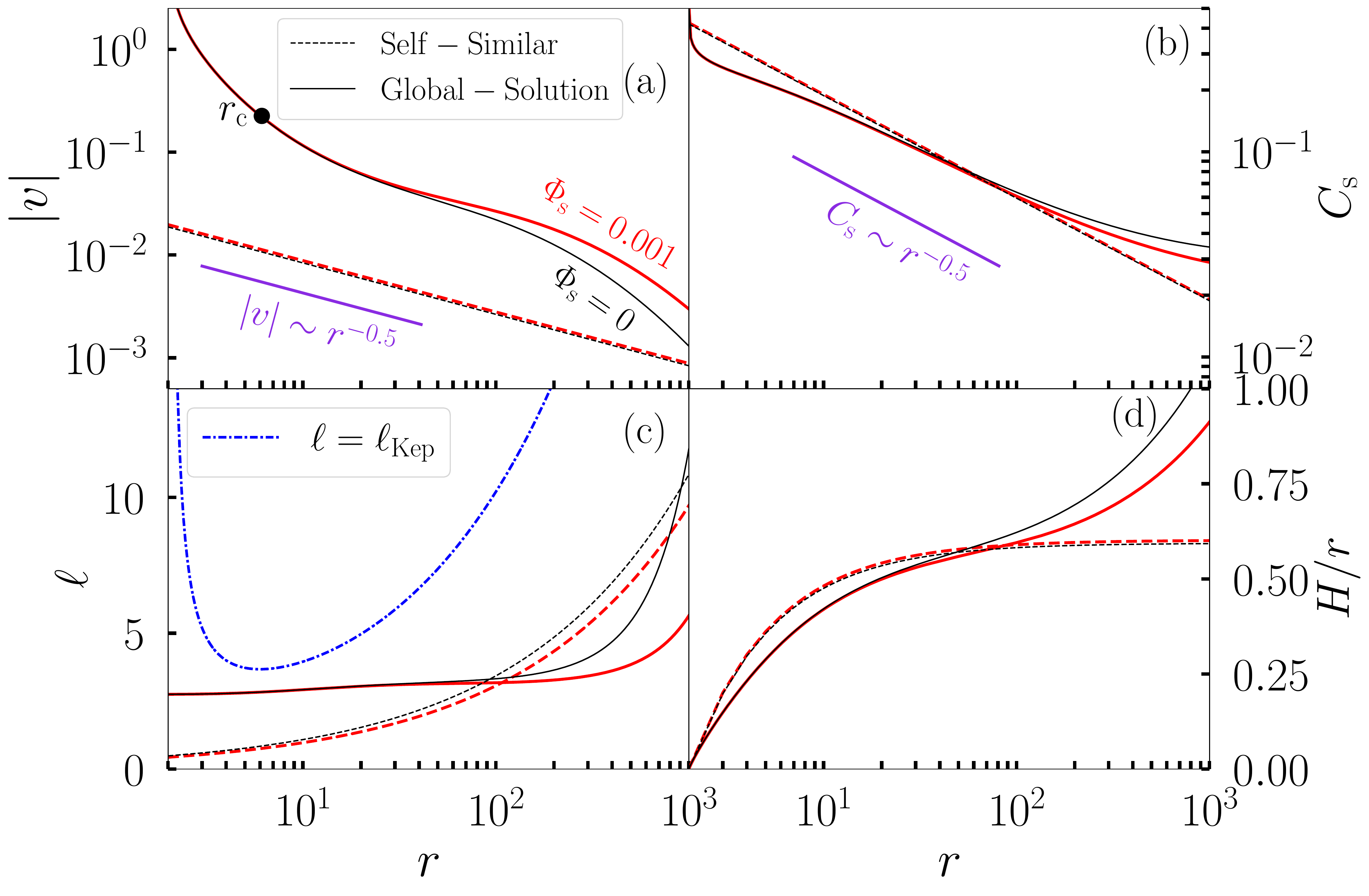}
	\caption{Comparison of global (solid curves) and self-similar (dashed curves) solutions in presence and absence of thermal conduction. In panels (a), (b), (c) and (d), radial velocity ($|v|$), sound speed ($C_{\rm s}$), angular momentum ($\ell$) and local disc thickness ($H/r$) are plotted. Here, we choose the input parameters for global solutions as $(\varepsilon_0,\ell_0,\alpha)=(0.0025,2.75,0.05) $ (same as in Fig. \ref{fig:variables}). The thin and thick curves represent results for $\Phi_{\rm s}=0$ and $\Phi_{\rm s}=0.001$, respectively. For self-similar solution, we choose $f=1$ and $\gamma=1.5$. See text for the details.
	 }
	\label{fig:Glob_Self}
\end{figure}

In Fig. \ref{fig:Glob_Self}, we compare the global transonic solutions with the self-similar solutions. While doing so, we choose the same set of input parameters for global solutions as used in Fig. \ref{fig:variables}, $i.e.$, $(\varepsilon_0,\ell_0,\alpha)=(0.0025,2.75,0.05)$. And, for self-similar solutions, we use $\alpha=0.05$, $f=1$, and $\gamma=1.5$, respectively. In panel (a), the profile of the radial velocity $|v|$ is presented and in panel (b), we show the variation of sound speed $C_{\rm s}$. For the self-similar solutions, the effects of thermal conduction in $ |v| $, and $C_{\rm s}$ appear to be insignificant even for high saturation constant $\Phi_s=0.001$. This happens because both radial velocity and sound speed follow simple power law as $ |v|, ~C_{\rm s} \sim r^{-1/2}$ having ignorable impact of thermal conduction. On contrary, the impact of $\Phi_{\rm s}$ is seen to be prominent on the global solutions. In addition, the Mach number $M~(=|v|/C_{\rm s})$ in global solutions generally decreases with radius, whereas it remains independent on $r$ for self-similar solutions \cite[]{Faghei2012mnras}. In fact, self-similar solutions do not possess critical point as they remain sub-sonic across the length scale of the disc. In panel (c), we illustrate the variation of angular momentum $\ell$ for the same solutions presented in Fig. \ref{fig:Glob_Self}(a). We find that $\ell$ is reduced with the increase of saturation constant $\Phi_{\rm s}$ particularly towards the outer part of the disc, which is in agreement with the results of \cite{Tanaka2006}. Moreover, $\ell$ steeply rises at larger radii as $\ell \propto r^{1/2}$ in self-similar approach, although HAFs remain sub-Keplerian all throughout provided $\alpha$ does not assume high end values. In panel (d), we present the variation of the local disc thickness $H/r$ as a function of $r$. We observe that in self-similar model, $H/r$ remains almost constant ($H/r \sim 0.6$) at the outer regions of the disc, whereas flow geometry becomes quasi-spherical ($H/r \sim 1$) for global solutions.

\begin{figure}
	\centering
	\includegraphics[width=\columnwidth]{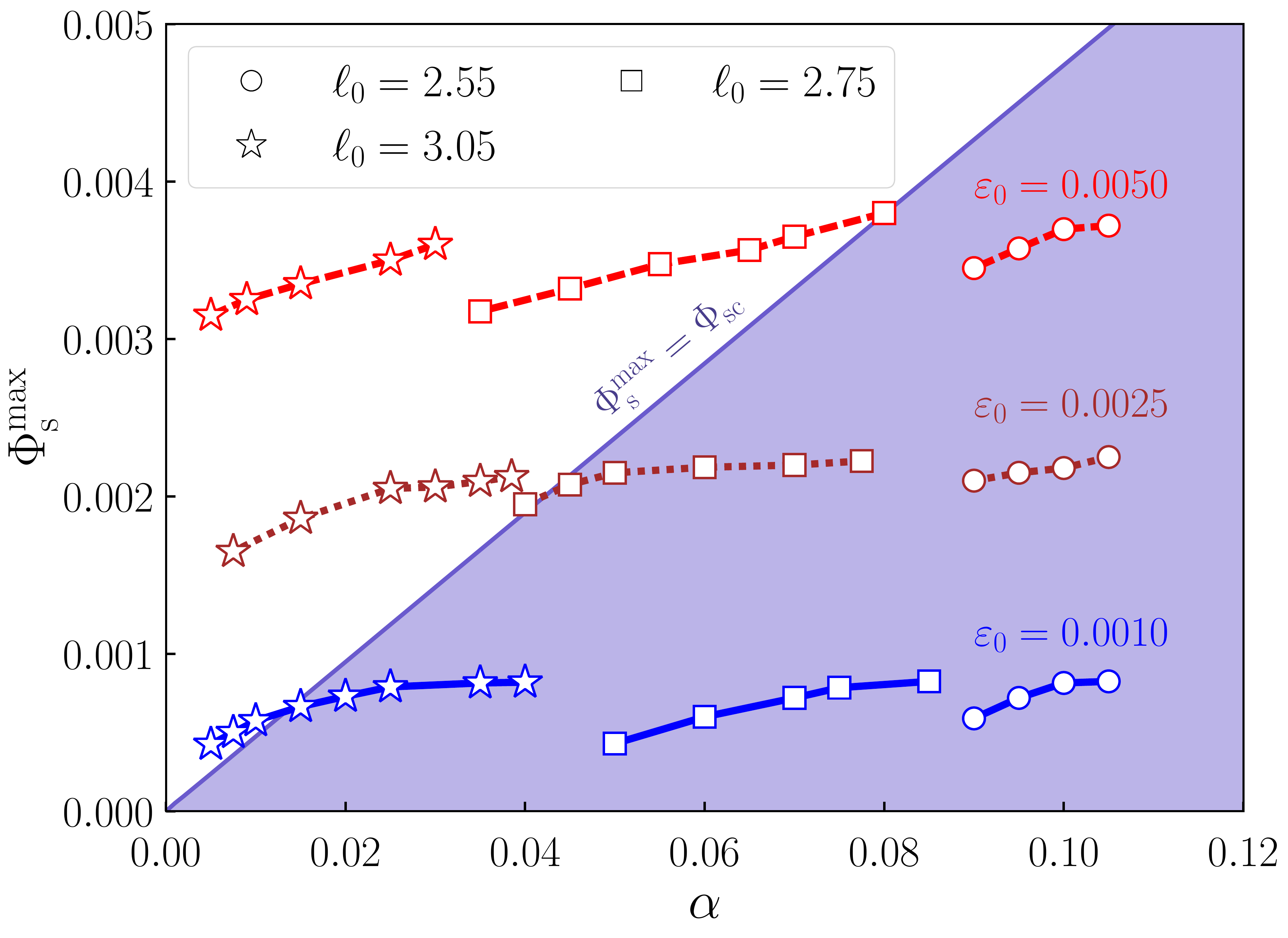}
	\caption{Correlation between $\alpha$ and maximum value of saturation constant $\Phi^{\rm max}_{\rm s} $ that renders the global transonic accretion solutions around black holes. Here, circles, squares, and asterisks are for $ \ell_0 = 2.55, 2.75$ and $3.05$ and these points connected with solid (blue), dotted (maroon), and dashed (red) lines represent the results for $ \varepsilon_0 =0.0010, 0.0025$, and $0.0050$, respectively. The shaded (violet) region corresponds to the self-similar solutions (see Eq. (\ref{eq:phialpha})) where slanting solid line refers the limiting value of saturation constant ($\Phi^{\rm max}_{\rm s}=\Phi_{\rm sc}$). See text for the details.		
	}
	\label{fig:alpha_phi}
\end{figure}

\subsection{Parameter space for global and self-similar solutions}
\label{sec:alpha-phi}

In this section, we put effort to determine the $\Phi_{\rm s}$ that admits global accretion solutions for a given set of input parameters $(\varepsilon_0, \ell_0, \alpha)$. Upon tuning the $(\varepsilon_0, \ell_0)$, we compute the maximum value of saturation constant $\Phi^{\rm max}_{\rm s} $ for a given $\alpha$ and present the obtain results in Fig. \ref{fig:alpha_phi}. Here, open circles, open squares, and open asterisks are for $ \ell_0 = 2.55, 2.75$ and $3.05$ and these points join using solid (blue), dotted (maroon), and dashed (red) lines corresponds to the results for $ \varepsilon_0 =0.0010, 0.0025$, and $0.0050$, respectively. We observe that for a set of $(\varepsilon_0, \ell_0)$, $\Phi^{\rm max}_{\rm s} $ increases with the increase of $\alpha$, which is in agreement with the results obtained from the self-similar solutions \cite[]{Ghasemnezhad2012,Faghei2012}. Further, we notice that for a given $\varepsilon_0$, when $\ell_0$ is small (high), flow with relatively higher (lower) viscosity admits global transonic solutions. On the other hand, for a given $\alpha$, when $\varepsilon_0$ is increased (decreased), the acceptable range of $\Phi^{\rm max}_{\rm sc}$ is also increased (decreased), irrespective to the choice of $\ell_0$ values. The shaded (violet) region corresponds to the self-similar solutions (see Eq. (\ref{eq:phialpha})) where slanting solid line refers the limiting value of saturation constant ($\Phi^{\rm max}_{\rm s}=\Phi_{\rm s}$). Here, we choose $f=1$, and $\gamma=1.5$. When $\Phi_{\rm s} > \Phi^{\rm max}_{\rm s} $, global solutions become infeasible and ceases to exist (see \S \ref{subsec:selfsimilar}). It is noteworthy that for lower $\ell_0$, $\Phi_{\rm sc}$ agrees well with $\Phi^{\rm max}_{\rm s}$ obtained from the global solutions. When $\ell_0$ is higher, a coarse agreement is observed for flows with lower $\varepsilon_0$ values. With this, we argue that the physically motivated global accretion solutions are prevalent than the simplistic self-similar solutions.

\begin{figure}
	\includegraphics[width=\columnwidth]{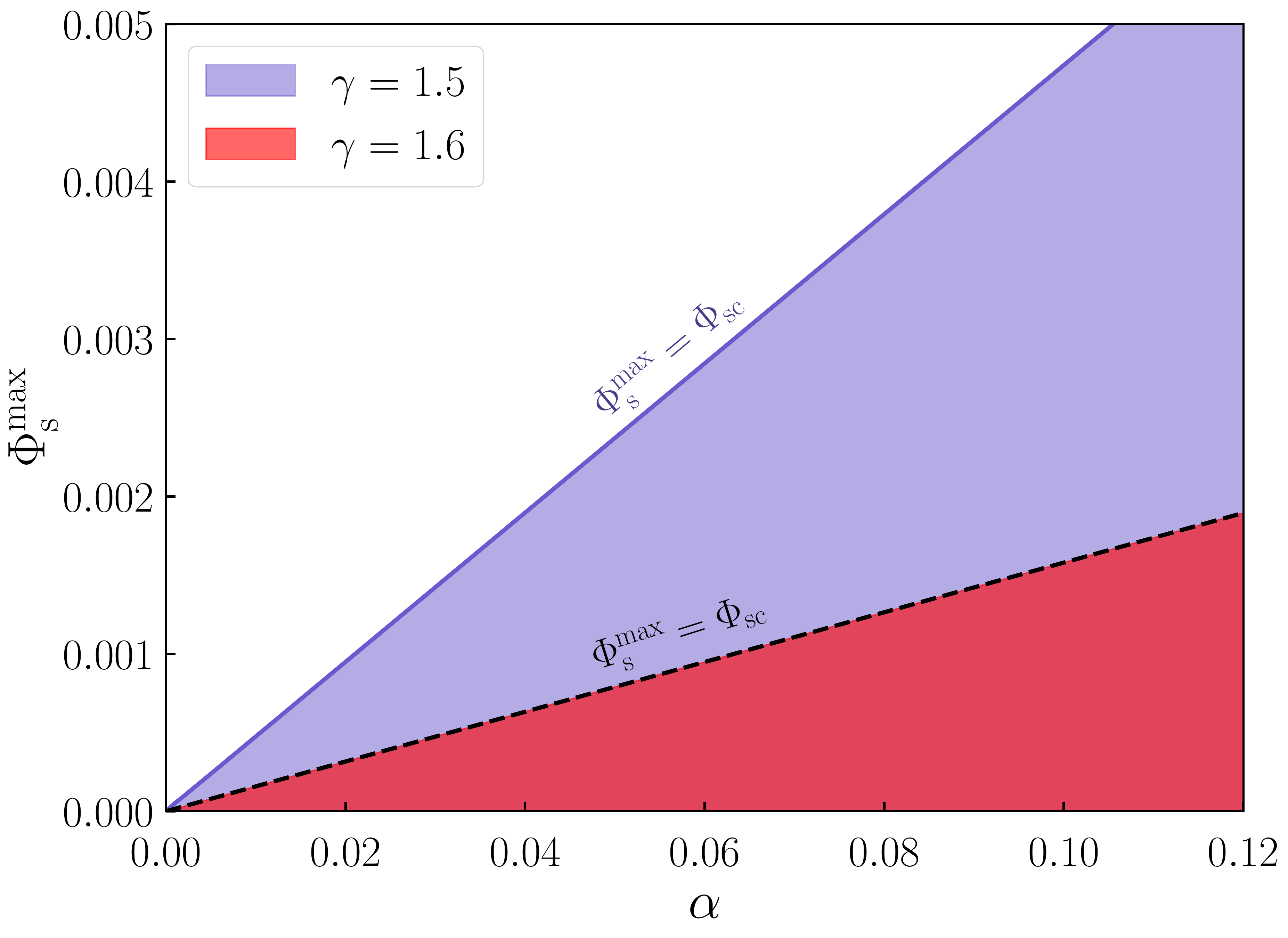}
	\caption{Correlation between $\alpha$ and $\Phi^{\rm max}_{\rm s} $ based on self-similar solutions (see equation (\ref{eq:phialpha})) for different adiabatic index $\gamma$. The shaded region in violet and red are for $\gamma=1.5$ and $1.6$, respectively. See text for the details.
	} \label{fig:f1}
\end{figure}

Next, we compare the limiting range of $\Phi^{\rm max}_{\rm s} = \Phi_{\rm sc}$ as a function of viscosity parameter $\alpha$ for different $\gamma$ values. The obtained results are shown in Fig. \ref{fig:f1}, where $\Phi^{\rm max}_{\rm s}$ is plotted as a function of $\alpha$. In the figure, the effective domain shaded in violet is for $\gamma=1.5$, whereas the same in red is obtained for $\gamma=1.6$. It is evident that the acceptable range of the saturation constant $\Phi_{\rm s}$ decreases as $\gamma$ is increased \cite[see also][]{Tanaka2006}. Based on this findings, we infer that self-similar solutions obtained using relatively lower $\gamma$ seems to be potentially more viable in articulating the features of global accretion solutions of HAFs (see Fig. \ref{fig:alpha_phi}).

\begin{figure}
	\includegraphics[width=\columnwidth]{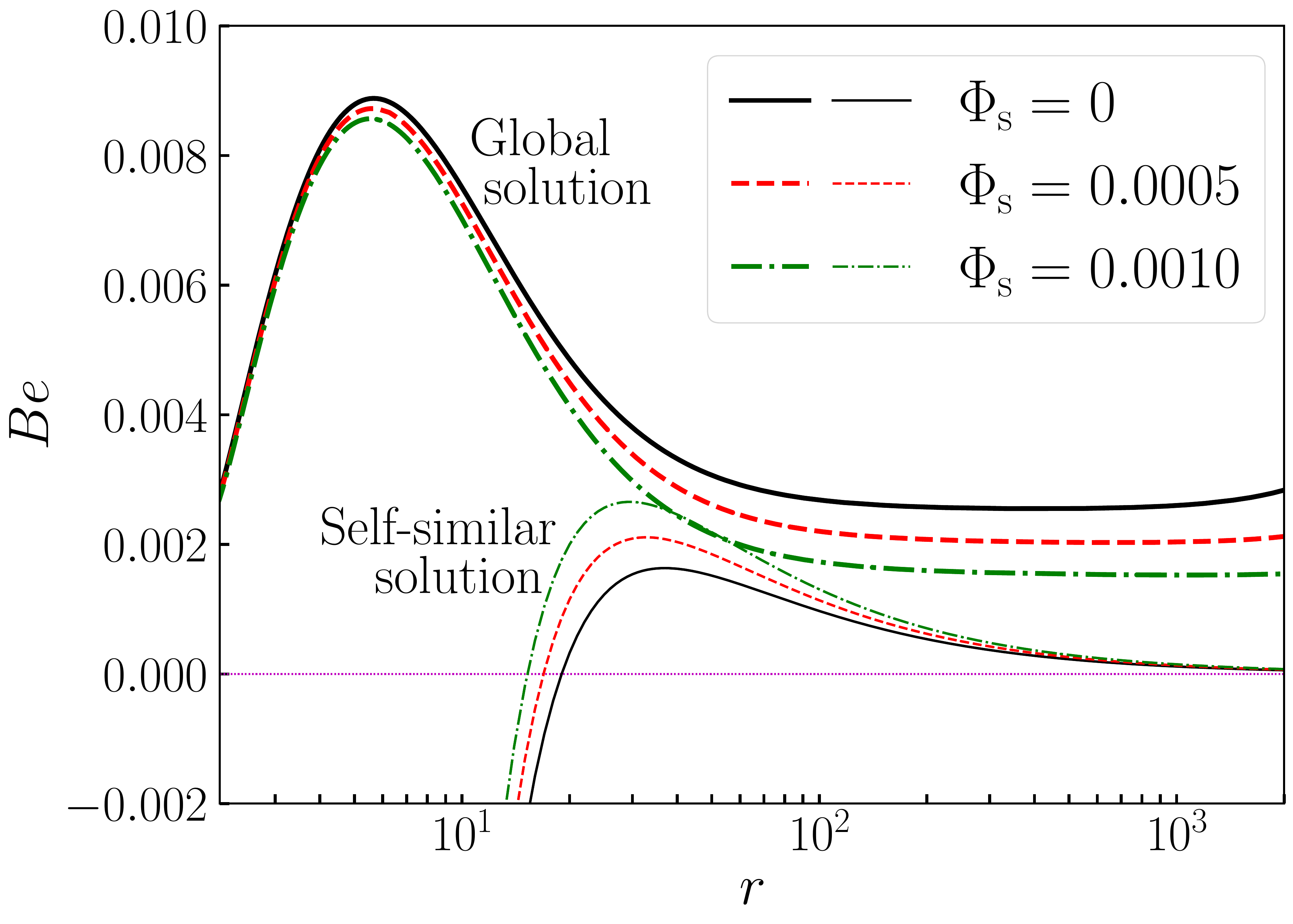}
	\caption{Plot of Bernoulli parameter ($Be$) as a function of logarithmic radial coordinate for different saturation constants $\Phi_{\rm s}$. Here, we choose the input parameters as $\gamma=1.5$, $f=1.0$, and $\alpha=0.05$, respectively. Thick curves represent results obtained from the global solutions using $\varepsilon_0=0.0025$, and $\ell_0 = 2.75$, whereas thin curves are for self-similar solutions. The solid (black), dashed (red) and dot-dashed (green) curves denote results for $\Phi_{\rm s}= 0, 0.0005$, and $0.0010$, respectively. The dotted (magenta) horizontal line corresponds to $Be=0$. See text for the details.
	}
	\label{fig:Be}
\end{figure}

\subsection{Bernoulli parameter}

In this section, we study the Bernoulli parameter $Be$ (see equation (\ref{eq:Be})) which coarsely accounts the evidence of outflow likely to be originated from the accretion disc. Accordingly, in Fig. \ref{fig:Be}, we display the typical variation of Bernoulli parameter $ Be $ as a function of radial coordinate ($r$). In the figure, thick curves correspond to the results obtained from the global accretion solutions, where input parameters are chosen as $\gamma=1.5$, $f=1.0$, $\alpha=0.05$, $\varepsilon_0 = 0.0025$, and $\ell_0 = 2.75$, respectively. Here, solid (black), dashed (red) and dot-dashed (green) curves represent results corresponding to $\Phi_{\rm s}=0.0, 0.0005$, and $0.0010$. Note that the overall profile of $Be$ is in agreement with the smooth solutions reported in \cite{Das-etal2009} \cite[see also][]{Kumar2018}. We find that the Bernoulli parameter $Be$ of global transonic solutions remain positive throughout the disc which is again in agreement with \cite{Narayan1997}. The positive Bernoulli parameter suggests that the accreting gas are unbound and therefore, a part of the accreting gas may escape (equivalently massloss) in the form of outflow with a net positive (kinetic) energy avoiding the strong gravitational pull of BH. With this, accreting gas tends to become energetically bound. However, these outflows are expected to be quite weak as the terminal Lorentz factor ranges $\Gamma = Be + 1 \sim 1.01$ \cite[]{Das-etal2009}. Indeed, shock-induced global accretion solutions seems potentially promising to generate powerful outflows having $\Gamma \sim 6$ \cite[]{Das-etal2009}, however, implementation of the shock physics is beyond the scope of the present paper and will be reported elsewhere.

It is worthy to compare the Bernoulli parameter obtained from global and self-similar solutions. Although the Bernoulli parameter of global transonic solution always remain positive, however, in self-similar approach, it often alters its sign from positive to negative as the accreting flow moves towards the black hole from the outer edge. In Fig. \ref{fig:Be}, we present the profile of $Be$ obtained from self-similar solutions using thin curves, where $\gamma=1.5$, $f=1.0$, $\alpha=0.05$ are used as input parameters. As before, the results plotted using solid (black), dashed (red) and dot-dashed (green) curves are for $\Phi_{\rm s}=0.0, 0.0005$, and $0.0010$, respectively. We observe that $Be$ becomes negative only at the inner part of the disc, where potential energy overcomes the remaining terms in equation (\ref{eq:Be}) yielding strongly bound flow. This happens because the radial and rotational velocities close to BH in the self-similar solutions are smaller than those in the global solutions. Further, we notice that the effect of thermal conduction on the Bernoulli parameter $Be$ is seen to be opposite. We infer that this finding possibly arises as the radial dependence of the disc variables in self-similar solutions (see equation (\ref{eq:v})$-$(\ref{eq:cs})) differs considerably from the global solutions when the thermal conduction is active inside the HAFs.

\section{SUMMARY AND DISCUSSION} \label{sec:summary_discussion}

In this paper, we present a comprehensive study of a low angular momentum, steady, axisymmetric, viscous, advective accretion flow around a non-rotating BH in presence of thermal conduction. Here, the conductive heat flux is described in the saturated form. This is because, the accretion flow becomes weakly collisional in such systems \cite[]{Quataert2004,Tanaka2006}. We adopt the pseudo-Newtonian potential introduced by \cite{Paczynsky1980} that satisfactorily mimics the space-time geometry around the non-rotating BHs. With this, we examine the effect of thermal conduction on the properties of the global transonic hot accretion flows around BHs.

The present model is based on the same set of governing equations that describe the advection dominated accretion flow (ADAF) \cite[]{Narayan1997}. Moreover, the conservation equations augmented by the inner boundary conditions \cite[]{Becker2003,Das-etal2009,Kumar2018} permit us to carry out the analysis from the location just out side the BH horizon $r_{\rm in}=2.001$. Using the model input parameters, namely energy ($\varepsilon_0$), angular momentum ($\ell_0$), viscosity parameter ($\alpha$), adiabatic index ($\gamma$), and saturation constant ($\Phi_{\rm s}$) and following the solution methodology presented in Appendix \ref{Method}, we obtain the complete set of global transonic solutions for the first time to the best of our knowledge in presence of thermal conduction. We summarize our findings below.

We find that the effect of thermal conduction on the global accretion solutions is significant particularly towards the outer part of the disc. When the saturation constant $\Phi_{\rm s}$ exceeds its limiting value, the nature of the global solution is altered and it becomes closed failing to connect the BH horizon with the outer edge of the disk (see Fig. \ref{fig:Mach_conduction1}). Solution of this kind remain unphysical unless it is connected via shock with another solution passing through a critical point usually located far from the horizon \cite[and referenes therein]{Fukue1987,Chakrabarti1996,Das-etal2001,Chakrabarti-Das2004,Becker2008,Das-etal2009,Das-etal2022}. Needless to mention that the studying shock-induced global accretion solution is beyond the scope of this paper and hence, will be reported elsewhere. Moreover, our results confirm that the thermal conduction affects the transonic properties of the HAFs. When $\Phi_{\rm s}$ is increased for flows with fixed input parameters ($\varepsilon_0,\ell_0,\alpha$), critical points recede away from the BH horizon (see Fig. \ref{fig:rc}).

We also examine the role of thermal conduction on the flow variables. We see that the increase of $\Phi_{\rm s}$ reduces the flow temperature, and disc height at the outer region (see Fig. \ref{fig:variables}). This possibly happens due to the fact that high thermal conduction generally weakens the viscous turbulence \cite[]{Faghei2012} that lowers the disc temperature. Indeed, this findings are in agreement with the results reported in \cite{Tanaka2006,Wu-etal2010}. Further, we compare the flow variables obtained by means of global and self-similar solutions and ample disagreement is seen (see Fig. \ref{fig:Glob_Self}). In fact, we observe that radial velocity and sound speed are not noticeably affected by thermal conduction for self-similar solutions. Notice that global accretion solutions remain sub-Keperian all throughout, however, self-similar solutions may become  super-Keplerian near the critical radius provided $\alpha$ assumes lower value \cite[]{Narayan1997,Chen1997,Kumar2018}.

One of the important results of this work is to identify the correlation between viscosity $\alpha$ and maximum saturation constant $\Phi^{\rm max}_{\rm s}$ that renders the global transonic solutions of HAFs. We find a positive correlation where $\Phi^{\rm max}_{\rm s}$ increases with $\alpha$ irrespective to the choice of ($\varepsilon_0,\ell_0$). We also observe that the flow with higher $\varepsilon_0$ can sustain higher $\Phi^{\rm max}_{\rm s}$ for global solutions, however, such dependencies are non-existence indicating the limitation of the self-similar approach (see Fig. \ref{fig:alpha_phi}).

In addition, we calculate the Bernoulli parameter $Be$ in the presence of thermal conduction to explore the possible existence of outflows in HAFs. Global solutions display a positive Bernoulli parameter at all radii, whereas self-similar solutions yield negative Bernoulli parameter at the inner part of the disc (see Fig. \ref{fig:Be}). Evidently, an accretion flow with positive Bernoulli parameter is unbound and therefore, matter is likely to escape from such unbound disc avoiding the strong gravity of BH in the from of outflow.

Finally, we mention the limitations of the present formalism as it is developed based on several approximations. We adopt pseudo potential to mimic the gravitational effect around a non-rotating black hole, instead of using general relativity. We consider single temperature disc assuming strong coupling existed between ion and electron. However, in HAFs, the ion-electron coupling generally becomes weak and hence, two-temperature flow structure seems to be viable at least at the inner part of the disc \cite[]{Rees1982,Yuan2014,Dihingia-etal2018,Dihingia-etal2020}. We neglect magnetic fields although the transport of angular momentum is expected due to the Maxwell stress associated with Magnetohydrodynamics (MHD) turbulence driven by magneto-rotational instability (MRI). Moreover, we refrain studying self-consistent accretion-ejection solutions that requires two dimensional approach. All these are indeed relevant, however, their implementations are beyond the scope of the present work. Indeed, we plan to take up these issues in our future works and will be reported elsewhere.

\section*{Acknowledgements}

Authors thank the anonymous reviewer for valuable comments and useful suggestions that help to improve the quality of the paper. SM acknowledges Prime Minister{}'s Research Fellowship (PMRF), Government of India for financial support. SM is indebted to Mr. Amit Kumar for valuable suggestions. AM is supported by the National Natural Science Foundation of China (grant No. 12150410308), foreign experts project (grant No. QN2022170006L), and the China Postdoctoral Science Foundation (grant No. 2020M673371). AM acknowledges the support of Dr. X. D. Zhang at the Network Information Center of Xi'an JiaoTong University. The computation work is done using the High Performance Computing (HPC) platform of Xi'an JiaoTong University. This work is supported by the Ferdowsi University of Mashhad under grant no. 57030 (1400/11/02). SD thanks Science and Engineering Research Board (SERB), India for support under grant MTR/2020/000331. SM and SD also thank the Department of Physics, IIT Guwahati, India for providing the facilities to complete this work. Authors acknowledge the Sci-HPC Center of the Ferdowsi University of Mashhad, where part of this research is performed. Authors also acknowledge the extensive use of the NASA Astrophysical Data System Abstract Service.

\section*{Data Availability}

The data underlying this article will be available with	reasonable request.
	
	
	
\bibliographystyle{mnras}
\bibliography{reference} 

\appendix
\section{Solution Methodology: Iteration Method} \label{Method}

We obtain the global transonic solutions using an iteration method. In this method, we begin the numerical integration of the flow equations from a location just outside the black hole horizon as $ r_{\rm in}=r_{\rm s} + 0.001 $. For a given set of input parameters $(\varepsilon_0, \ell_0,\alpha, \Phi_{\rm s})$, we compute three flow variables, namely velocity $ v_{\rm in} $, sound speed $ C_{{\rm s}_{\rm in}}$, and angular momentum $\ell_{\rm in}$ at $ r_{\rm in} $. Needless to mention that $ 0 < \ell_0 < \ell_{\rm ms} $, where $ \ell_{\rm ms} = 2\sqrt{3}$ is the innermost stable angular momentum around a Schwarzschild BH. Considering this, we pursue the following chronology for obtaining the global solutions.

\begin{itemize}
	
\item Close to horizon \text{i.e.,} $r - r_{\rm s} \rightarrow{0} $, the matter falls with free fall velocity $ v_{\rm ff} = -\sqrt{2 / (r - r_{\rm s})}$ . Here, we consider a fractional constant $ \delta < 1$, such that $ v_{\rm in} = \delta \times v_{\rm ff}$, and accreting matter enters into the BH with this velocity $ v_{\rm in}$.

\item Next, we determine the asymptotic flow variables just outside the horizon at $ r_{\rm in} $. Using Frobenius expansion, we get the asymptotic behaviour of flow angular momentum as,
\begin{equation}
        \ell_{\rm in} = \ell_0 + B (r - r_{\rm s})^\beta; \hspace{1cm}  r\rightarrow r_{\rm s},
\end{equation}
where $B$ and $\beta$ are positive constants. We implement $\ell_{\rm in}$ in equation (\ref{eq:domg}) and we demand that,
\begin{equation}\label{l_asymp}
        \lim_{r \to r_{\rm s}} \frac{d\ell_{\rm in}}{dr} = 0,\hspace{1cm}
        \lim_{r\to r_{\rm s}} \frac{\delta \sqrt{2} B (r-r_{\rm s})^\beta}{\sqrt{r_{\rm s}} (r-r_{\rm s})^\frac{3}{2} \alpha C_{\rm s}^2} = \frac{2 \ell_0}{r_{\rm s}}.
\end{equation}
To eliminate all terms involving $(r-r_{\rm s})$ in equation (A2), we require $\beta = 3/ 2$ and $B = \sqrt{2/r_{\rm s}} (\alpha C_{\rm s}^2 \ell_0 / \delta)$. Accordingly, we get the expression of angular momentum at $r_{\rm in}$ for a suitable choice of $(\delta,\ell_0,\alpha,\Phi_{\rm s})$.

\item We use $v_{\rm in}$ and $\ell_{\rm in}$ in equation (\ref{eq:conservedE}), where we consider that the effect of conduction is negligible \cite[]{Tanaka2006}, and determine the sound speed $C_{s_{\rm in}}$ by solving equation (\ref{eq:Econs}) for a given energy $\varepsilon_0$,

\item Using $v_{\rm in}$, $ C_{{\rm s}_{\rm in}} $, and $\ell_{\rm in}$, we start integrating equations (\ref{eq:domg}, \ref{eq:dcs}, \ref{eq:dv}) from $r_{\rm in}$ outwards, and check the critical point conditions described in equations (\ref{eq:D}) and (\ref{eq:N}). We keep tuning the iteration parameter $ \delta $ until the critical point conditions are satisfied for $ \delta = \delta_{\rm c} $, and thereafter, we obtain the critical point $ r_{\rm c} $ (see Fig. \ref{fig:method_delta} for more details).

\begin{figure}
	\centering
	\includegraphics[width=\columnwidth]{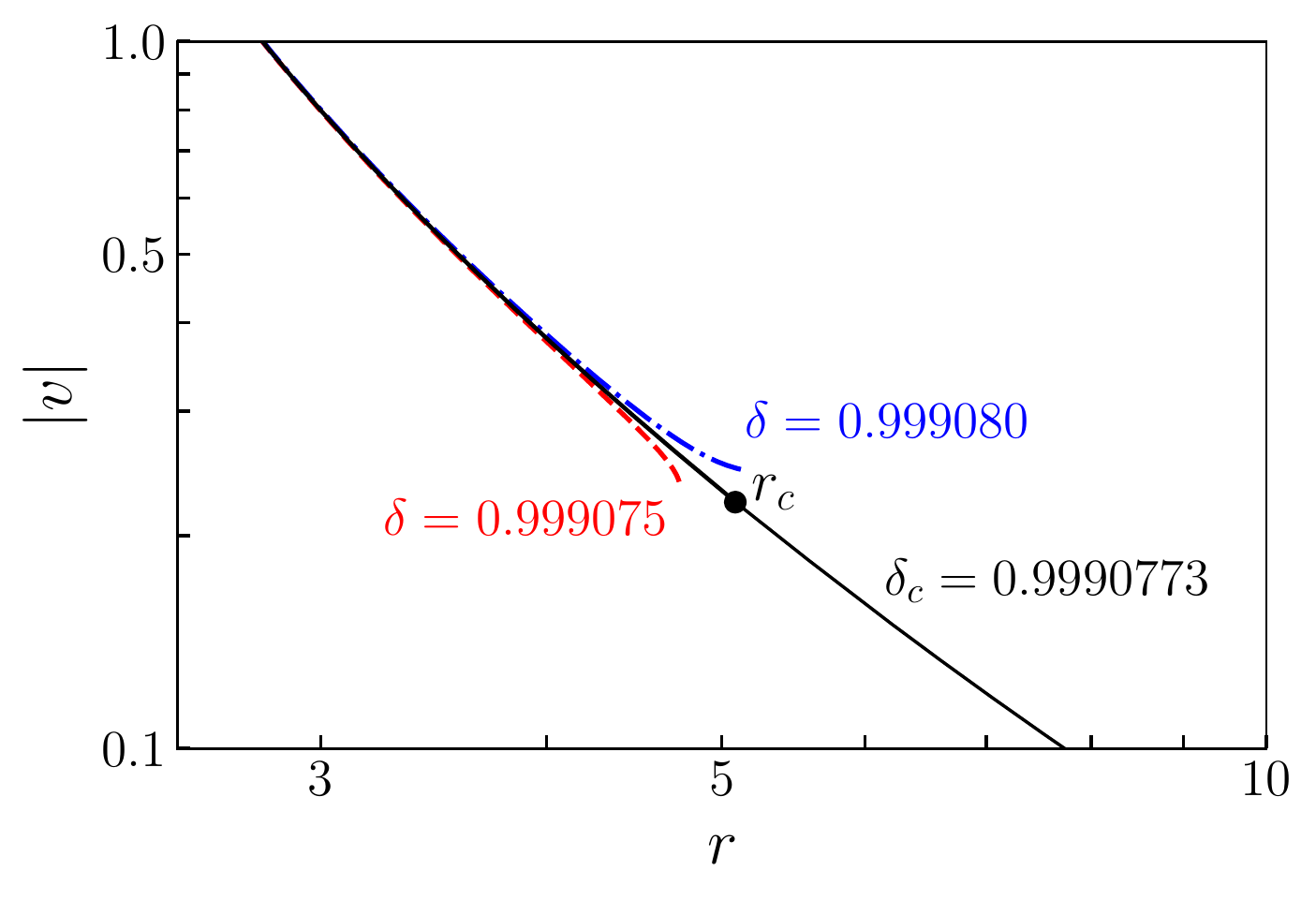}
	\caption{Variation of flow velocity $ |v| $ as a function of radial coordinate $ r $ for three different iteration parameters. Dashed (red), solid (black) and dot-dashed (blue) curves denote results for $\delta=0.999075, 0.9990773$, and $0.999080$. Here, $\delta=\delta_c=0.9990773$ corresponds to transonic solution where critical point is obtained at $ r_{\rm c} = 5.0286 $ for the chosen input parameters ($ \varepsilon_0, \ell_0, \alpha, \Phi_{\rm s}, \gamma )=( 0.005, 3.05, 0.04, 0.0015, 1.5 $).
	}
	\label{fig:method_delta}
\end{figure}

\item At $ r_{\rm c} $, we calculate $dv/dr|_{\rm c}$ by applying the l$'$H\^{o}pital's rule. The real and negative radial velocity gradient corresponds to accretion solution, and hence, for $ dv/dr|_c < 0 $, we further integrate equations (\ref{eq:domg}, \ref{eq:dcs}, \ref{eq:dv}) starting from $ r_{\rm c} $ upto to the outer edge of the disc $ r_{\rm edge}=1000$. Finally, we join both parts of the solutions (from $r_{\rm in}$ to $r_{\rm c}$ and $r_{\rm c}$ to $ r_{\rm edge}$) to obtain the global transonic accretion solution for a HAF around non-rotating BH.

\end{itemize}

\bsp	
\label{lastpage}
\end{document}